\input{aipcheck}

\documentclass[
    ,final            
  ]
  {aipproc}

\usepackage{psfig}
\usepackage{graphicx}
\usepackage{epsfig}
\usepackage{subfigure}

\layoutstyle{6x9}

\begin{document}

\title{Factorization of Short- and Long-range Interactions in Charged Meson Production}


\author{Tanja Horn}{
  address={Physics Division, TJNAF, Newport News, Virginia, 23606}
}

\begin{abstract}
Meson production data play an important role in our understanding of nucleon
structure. The combination of reaction channels is sensitive to gluon and charge and flavor non-singlet quark densities and has the potential to provide detailed images of the QCD quark structure of the nucleon. Quark imaging requires a good understanding of the reaction mechanism, and in particular rigorous tests of factorization of long- and short-distance physics. The higher energies after the Jefferson Lab 12 GeV upgrade provide ideal conditions for such studies, which are an essential prerequisite for studies of valence quark distributions. An electron-ion collider would allow to extend these studies to detailed imaging of sea quarks and gluons. 
\end{abstract}

\maketitle


Depending on the energy, the meson production reaction mechanism can be viewed in two different ways. At low virtualities of the photon, $Q^2$, the process can be described as the hadronic fluctuations of the virtual photon and their interaction with the target. The system is characterized by the invariant mass of the photon-nucleon system, $W$, and the four-momentum transfer to the nucleon $-t$. In the limit of small $-t$ and $W$, the process is dominated by the $t$-channel exchange meson pole term, which is dominated by longitudinally polarized photons. The dominance of these longitudinal contributions allows for extracting the meson form factor which describes the spatial distribution of quarks in the nucleon. At sufficiently high values of $Q^2$, the process becomes effectively pointlike, and can be understood as the interaction of the virtual photon with partonic degrees of freedom. This process is presented schematically in terms of the ``handbag'' diagram, which consists of a perturbative (hard scattering) part (exchange of a single hard gluon) and a non-perturbative (soft) part that is represented by the Generalized Parton Distributions (GPDs). A QCD factorization theorem~\cite{collins97}, proven for longitudinal photons, allows one to compute the electroproduction amplitude in terms of the GPDs.

GPDs and meson form factors are essential to our understanding of the structure of hadrons. An advantage to use meson electroproduction is that is allows one to study both GPDs and meson form factors, shedding light on the quark-antiquark interaction in QCD. Before one can study meson form factors, however, one has to make sure that the longitudinal cross section, $\sigma_L$, is dominated by the $t$-channel meson exchange (``meson pole'') at small values of $-t$. Similarily, before one can extract any physical information contained in GPDs, one has to demonstrate that the factorization of the GPD from the hard scattering process is indeed applicable. 

\begin{figure}
\includegraphics[height=0.3\textheight]{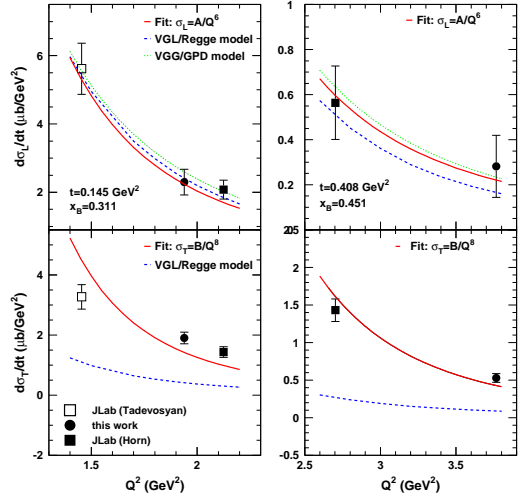}
\caption{The $Q^2$ dependence of the L/T separated $\pi^+$ cross section.}
\label{fig:xsec}
\end{figure}

Extractions of GPDs from experimental data are limited to the kinematic regime where factorization of long- (soft) and short-distance (hard) physics applies. One of the most stringent tests is the $Q^2$ dependence of the longitudinal cross section. The meson QCD factorization theorem predicts that $\sigma_L$ scale like $Q^{-6}$, while transverse contributions are expected to be suppressed by $1/Q$ in the amplitude. As a consequence, the longitudinal cross section is expected to dominate at very large values of $Q^2$. Figure~\ref{fig:xsec} illustrates the $Q^2$ dependence of L/T separated pion electroproduction cross sections from Jefferson Lab~\cite{horn06,blok08,horn08}. The solid red line represents a fit to the data and the dashed lines theoretical predictions from VGL/Regge~\cite{vgl} and VGG/GPD~\cite{vgg} calculations. The QCD scaling prediction is reasonably consistent with these recent $\pi^+$ $\sigma_L$ data, but $\sigma_T$ does not follow the scaling expectation. The longitudinal cross sections are reasonably well described by non-perturbative Regge calculations and the calculations performed in the GPD framework. The transverse cross section is significantly larger than the longitudinal cross section at $Q^2>$ 2 GeV$^2$ and also the theoretical prediction. 

The $Q^2$ dependence of the meson form factor provides another perspective on soft-hard factorization. The prediction in the very high $Q^2$ regime is $1/Q^2$. Current precision pion form factor data up to $Q^2$=2.45 GeV$^2$~\cite{horn06,horn08} seem to follow the prediction from perturbative QCD (pQCD) and seem to suggest that the factorization prediction holds. However, against expectation for perturbative behavior the magnitude of the form factor is large. This could, for instance, be due to the fact that the factorization prediction does not hold or that something is missing in the hard calculation. This behavior is similar to the one observed in the cross section, where $\sigma_L$ and the form factor scale as predicted, $\sigma_T$ and the magnitude of the form factor are large against expectation.

There are at least two approaches to investigate this intriguing behavior. The first is to extend the kinematic reach of the current charged pion data. Approved Jefferson Lab experiment E12-07-105~\cite{E12-07-105} will use the new focusing spectrometer in Hall C (SHMS) to search for the onset of factorization in pion production. This will extend the current 6 GeV $Q^2$ coverage by 2-3 times at smaller values of $-t$. This experiment will be essential for a reliable interpretation of results from the Jefferson Lab GPD program at both 6 GeV and 12 GeV. The experiment will address questions such as if the partonic description is applicable in practice and if it is possible to extract GPDs using pion production. L/T separations are essential for these studies. In particular, one needs access to $\sigma_L$ from which one extracts the GPD. If $\sigma_L$ in meson production is small, GPD flavor studies may be limited to focusing spectrometers. However, recent data suggest that $\sigma_L$ in $\pi^-$ production is larger than for $\pi^+$. If this holds, one could extract $\sigma_L$ from the unseparated cross section. In addition to extending the kinematic reach E12-07-105 will thus also compare $\pi^+$ and $\pi^-$ production to check possibilities of extracting GPDs without explicit L/T separation. While the observed trends in $\pi^-$ production are promising, $\sigma_T$ in $\pi^+$ production is much larger than theoretical predictions from the VGL/Regge model (see Fig.~\ref{fig:xsec}). More recent calculations~\cite{giessen,obukhovsky} provide a somewhat better description. To understand the reaction mechanism however, it would be beneficial to compare with a different yet similar system. The best candidate for such studies is the kaon. Data taken in the resonance region~\cite{mohring03} suggest that $\sigma_T$ for  $K^+$ production is also not small even at $Q^2$=2 GeV$^2$. Unfortunately, available kaon data are limited. There are no separated data above the resonance region, and the existing data have large uncertainties~\cite{Beb74}. 

As a second approach to investigate current curious observations in meson production one can take advantage of the higher energies of the Jefferson Lab 12 GeV upgrade and vary the system. Approved experiment E12-09-011~\cite{E12-09-011} will provide the first L/T precision separated exclusive charged kaon production data above the resonance region ($W>$2 GeV). It includes high-$Q^2$ measurements, which allow for a quasi model independent comparison of kaon and pion production in the scaling regime, which is believed to be more robust than calculations of absolute cross sections based on QCD factorization and present GPD models. As mentioned above, in $\pi^+$ production, $\sigma_L$ at low $-t$ is well described  by Reggeized Born term models and found to be dominated by the pion pole term. Regarding $\sigma_T$ there are also intriguing suggestions from  recent theoretical studies of Jefferson Lab 6 GeV results that the reaction mechanism in exclusive pion production could be described as the limiting case of semi-inclusive production~\cite{giessen}. This is another strong argument for investigating the behavior of $\sigma_T$ in the kaon system.

To reliably extract the meson form factor, the influence of the non-pole $t$-channel contributions have to be modest in comparison to pole contributions. In practice this means that the ratio $\sigma_L$/$\sigma_T$ has to be large. For kaons, our current knowledge of $\sigma_L$ and $\sigma_T$ above the resonance region is insufficient. Experiment E12-09-011 will enhance our understanding of the $t$ channel kaon exchange in the amplitude. If the pole term in $\sigma_L$ indeed dominates, these data may be used to extract the kaon form factor. If one can extract the kaon form factor, it would be interesting to revisit the meson form factor puzzle. By comparing the observed $Q^2$ dependence and magnitude of the $\pi^+$ and $K^+$ form factors one could test if the analogy between pion cross section and form factor will also manifest itself for kaons. Indeed, these two data sets together will address the question if scaling is different for kaons than pions, and also allow for a quasi-model independent studies in the regime towards the onset of factorization.

A possible next step beyond Jefferson Lab at 12 GeV, where one can image the 
valence quarks of the nucleon, is to explore the sea of virtual quarks and gluons. These studies could be carried out at an Electron-Ion Collider (EIC), which is a next-generation facility recommended in the NSAC Long Range Plan in 2007. It could provide unprecedented access to gluon imaging in nucleons and nuclei. In general, the two possible physics goals that have been suggested for the EIC are 1) Studies of QCD at high gluon densities, and 2) Studies of nucleon structure beyond the valence region, which includes precision imaging of sea-quarks and gluons to determine spin, flavor, and spatial structure of the nucleon. The candidates for building the EIC are BNL and Jefferson Lab. Regarding meson production, the interest lies in the comparison of the different reaction channels. The latter can be divided into two categories. The first one contains the ``diffractive'' channels, \textit{e.g.}, $\rho^0$ and $J/\Psi$, which provide access to gluon imaging of the nucleon. The second one contains the ``non-diffractive'' processes, \textit{e.g.}, $\pi^+$, $\pi^0$, and $K \Lambda$, which are sensitive to non-singlet quarks and provide information about the spin and flavor of quark GPDs. Measurements of exclusive processes require high luminosity and the capability to detect the recoiling baryon at small angles. For the detector design at the EIC it is thus important to have a good understanding of the kinematics of the reactions. Furthermore, rate studies are important to check the feasibility of measuring a particular channel. In both cases simulation studies by Horn \textit{et al.} have shown that exclusive reactions have the most stringent requirements and thus drive the detector design. 

In summary, measurements of exclusive meson production at Jefferson Lab and a future EIC have the potential to contribute significantly to our understanding of nucleon structure.


\begin{theacknowledgments}
This work was supported in part by the U.S. Department of Energy. The 
Southeastern Universities Research Association (SURA) operates the Thomas
Jefferson National Accelerator Facility for the United States Department of
Energy under contract DE-AC05-84150.
\end{theacknowledgments}

\end{document}